\documentclass[pre,a4paper,showpacs,preprint,byrevtex]{revtex4}
\usepackage{graphicx}
\usepackage{dcolumn}
\usepackage{amsmath}
\usepackage{bm}
\usepackage{array}

\begin{document}

%\preprint{Preprint Universit\'{e} de Mons-Hainaut}
%\draft
\title{Lagrange mesh, relativistic flux tube, and rotating string}

\author{Fabien \surname{Buisseret}}
\thanks{FNRS Research Fellow}
\email[E-mail: ]{fabien.buisseret@umh.ac.be}
\author{Claude \surname{Semay}}
\thanks{FNRS Research Associate}
\email[E-mail: ]{claude.semay@umh.ac.be}
\affiliation{Groupe de Physique Nucl\'{e}aire Th\'{e}orique,
Universit\'{e} de Mons-Hainaut,
Acad\'{e}mie universitaire Wallonie-Bruxelles,
Place du Parc 20, B-7000 Mons, Belgium}

\date{\today}

\begin{abstract}
The Lagrange mesh method is a very accurate and simple procedure to
compute eigenvalues and eigenfunctions of nonrelativistic and
semirelativistic Hamiltonians. We show here that it can be used
successfully to solve the equations of both the relativistic flux tube
model and the rotating string model, in the symmetric case.
Verifications of the convergence of the method are given.
\end{abstract}

% PACS
% 02.70.-c Computational techniques
% 03.65.Ge Solutions of wave equations: bound states
% 12.39.Ki Relativistic quark model
% 02.30.Mv Approximations and expansions

\maketitle

\section{Introduction}
\label{Intr}

The Lagrange mesh method is a very accurate and simple procedure to
compute eigenvalues and eigenfunctions of a two-body Schr\"{o}dinger
equation \cite{baye86,vinc93,baye95}. The trial eigenstates are
developed in a basis of well-chosen functions, the Lagrange functions,
and the Hamiltonian matrix elements are obtained with a Gauss
quadrature. This method can be extended to treat very accurately
three-body problems, in nuclear or atomic physics \cite{hess99}.
Recently, it has also been successfully applied to a two-body spinless
Salpeter equation \cite{sem01}. The idea of this work is to adapt the
Lagrange mesh method to solve the complicated equations of both the
relativistic flux tube and the rotating string models.

\par The relativistic flux tube (RFT) is a phenomenological model
describing the mesons. It relies on the assumption that the quark and
the antiquark are connected by a straight color flux tube carrying both
energy and momentum. The quarks are considered as spinless particles in
the original version of the model \cite{laco89,olss95,sema95}. The RFT
reproduces the linear Regge trajectories, and reduces to the usual
Schr\"{o}dinger equation with a linear confinement potential in the
nonrelativistic limit. We will consider here the particular case of
mesons composed of two equal quark masses. The equations of motion of
the symmetric RFT model are given by two coupled nonlinear equations:
one defining the Hamiltonian and the other defining the orbital angular
momentum. These equations depend on a quark transverse velocity operator
and their solutions will be obtained by the use of an iterative
procedure similar to the one proposed in Ref.~\cite{sema95}.

\par The rotating string model (RS) also describes the mesons. It is
derived from the QCD Lagrangian and is characterized by the fact that it
contains auxiliary fields. The equations of motion for this model are
similar to the equations of motion of the RFT model
\cite{dubi94,morg99}. In the symmetric case, it has been showed that the
RS is classically equivalent to the RFT if the auxiliary fields are
correctly eliminated \cite{sema04}. This result, extended recently to
the asymmetric case \cite{buis04}, provides a clear physical
interpretation for the characteristic variables of the RS model.

\par The Lagrange mesh method is explained in Sec.~\ref{Meth}.
In Sec.~\ref{models}, the relativistic flux tube and the rotating string
models are described. Then, it is shown, in Sec.~\ref{resolution}, how
the Lagrange mesh method can be applied to solve the equations of motion
of these models. After some remarks, given in Sec.~\ref{params}, about
the numerical and physical parameters, the results are presented in
Sec.~\ref{Resu} and the reliability of our numerical method is checked.
Finally, some concluding remarks are given in Sec.~\ref{conclu}.

\section{Lagrange mesh method}\label{Meth}

A Lagrange mesh is formed on $N$ mesh points $x_{i}$ associated with an
orthonormal set of indefinitely derivable functions $f_{j}(x)$
\cite{baye86,vinc93,baye95}. A Lagrange function $f_{j}(x)$
vanishes
at all mesh points but one; it satisfies the Lagrange conditions
\begin{equation}
\label{flagpro}
f_{j}(x_{i})=\lambda^{-1/2}_{i}\delta_{ij}.
\end{equation}
The mesh points $x_{i}$, the zeros of a particular polynomial, and the
$\lambda_{i}$ are connected with a gauss quadrature formula
\begin{equation}
\label{gauss}
\int^{b}_{a} g(x)\, dx  \approx\sum^{N}_{k=1}\lambda_{k}\, g(x_{k}),
\end{equation}
used to compute all the integrals over the interval $[a,b]$.

As we consider only radial equations, this interval is $[0,\infty[$,
leading
to a Gauss-Laguerre quadrature. The Gauss formula (\ref{gauss}) is
exact when $g(x)$ is a polynomial of degree $2N-1$ at most, multiplied
by $\exp(-x)$. The Lagrange-Laguerre mesh is then based on the zeros of
the Laguerre polynomial $L_{N}(x)$ of degree $N$ \cite{baye86}.
An explicit form can be derived for the corresponding regularized
Lagrange functions
\begin{equation}
\label{flag}
f_{i}(x)=(-1)^{i}x^{-1/2}_{i}\, x(x-x_{i})^{-1}L_{N}(x)\, e^{-x/2}.
\end{equation}

\par To show how these elements can be applied to a physical problem,
let us consider a Hamiltonian $H = T(\vec{p}^{\, 2})+V(r)$, where
$T(\vec{p}^{\, 2})$ is the kinetic term and $V(r)$ a radial potential
($\hbar = c=1$).
The calculations are performed with trial states $|\psi\rangle$ given by
\begin{equation}
\label{state}
\left|\psi\right\rangle=\sum^{N}_{k=1}C_{k}\left|f_{k}\right\rangle,
\end{equation}
where
\begin{equation}
\left\langle \vec{r}\,|f_{k}\right\rangle=
\frac{f_{k}(r/h)}{\sqrt{h}r}Y_{\ell m}(\hat{r}).
%\equiv R(r)Y_{\ell m}(\hat{r}).
\end{equation}
$\ell$ is the orbital angular momentum quantum number and the
coefficients $C_{k}$ are
linear variational parameters. $h$ is the scale parameter chosen to
adjust the mesh to the domain of physical interest. We define $r=h\,x$,
with $x$ a dimensionless variable.

\par We have now to compute the Hamiltonian matrix elements. Using the
properties of the Lagrange functions and the Gauss
quadrature~(\ref{gauss}), the potential matrix is diagonal.
Its elements are
\begin{equation}
\label{poten}
\left\langle f_{i}|V(r)|f_{j}\right\rangle\approx
V(hx_{i})\delta_{ij},
\end{equation}
and only involve the value of the potential at the mesh points.
As the matrix elements are computed only approximately, the variational
character of the method cannot be guaranteed. But the accuracy of the
method is preserved \cite{baye02}.

The kinetic energy operator is only a function of $\vec{p}^{\, 2}$. Let
us define the corresponding matrix,
\begin{equation}
P^{2}_{ij}=\left\langle f_{i}|\vec{p}^{\, 2}|f_{j}\right\rangle.
\end{equation}
It is shown in Ref.~\cite{baye95} that, using the Gauss quadrature and
the properties of the Lagrange functions, one obtains
\begin{equation}
P^{\, 2}_{ij}=\frac{1}{h^{2}}\left(p^{\, 2}_{r\,
ij}+\frac{\ell(\ell+1)}{x^{2}_{i}}\delta_{ij}\right),
\end{equation}
where
\begin{equation}\label{pij_def}
p^{2}_{r\, ij}=\left\{
\begin{array}{lll}
&(-1)^{i-j}(x_{i}x_{j})^{-1/2}(x_{i}+x_{j})(x_{i}-x_{j})^{-2} &(i\neq
j),\\
&(12x^{2}_{i})^{-1}[4+(4N+2)x_{i}-x^{2}_{i}]&(i=j).
\end{array} \right.
\end{equation}
Now, the kinetic energy matrix $T(P^{2})$ can be computed with the
following method \cite{sem01}:
\begin{enumerate}
\item Diagonalization of the matrix $P^{2}$. If $D^{2}$ is the
corresponding diagonal matrix, we have
\begin{equation}
P^{2}=SD^{2}S^{-1},
\end{equation}
where $S$ is the transformation matrix.
\item Computation of $T(D^{2})$ by taking the function $T$ of all
diagonal elements of $D^{2}$.
\item Determination of the matrix elements
$T_{ij}=\left\langle f_{i}|T(P^{\, 2})|f_{j}\right\rangle$ in the
Lagrange basis by using the transformation matrix $S$
\begin{equation}
T(P^{\, 2})=ST(D^{2})S^{-1}.
\end{equation}
\end{enumerate}
This procedure can easily be generalized to the case of an arbitrary
function $F$ of any given matrix $M$, in order to compute $F(M)$
(provided the calculation is relevant). Note
that such a calculation is not exact because the number of Lagrange
functions is finite. However, it has already given good results in the
semirelativistic case, where
$T(\vec{p}^{\, 2})=\sqrt{\vec{p}^{\, 2}+m^{2}}$ \cite{sem01}.

\par The eigenvalue equation reduces to a system of $N$ mesh equations
\begin{equation}
\sum^{N}_{j=1}\left[ T_{ij}+V(hx_{i})\delta_{ij}-E \delta_{ij}\right]
C_{j}=0 \quad \text{with} \quad C_{j}=\sqrt{h\lambda_{j}}u(hx_{j}),
\end{equation}
where $u(r)$ is the regularized radial wave function. The coefficients
$C_{j}$ provide the values of the radial wave function at mesh points.
But contrary to some other mesh methods, the wave function is also known
everywhere thanks to Eq.~(\ref{state}).

\section{The models}
\label{models}

\subsection{The relativistic flux tube}
\label{Rela}

In the original RFT model \cite{laco89}, the meson is composed by two
spinless particles - a quark and an antiquark - which move being
attached with a flux tube. This tube is assumed to be linear with a
uniform constant energy density $a$ and carries angular momentum. A tube
element has only a transverse velocity. The system rotates in a plane
with a constant angular velocity $\omega$ around the center of mass,
assumed to be stationary. If $r_{i}$ is the distance between the $i$th
quark and the center of mass, and if we define $\dot{r}_{i}=dr_{i}/dt$
the radial velocity oh the $i$th quark, then the quark speed is given by
$v^{2}_{i}=\dot{r}^{2}_{i}+v^{2}_{i\bot}$, where
$v_{i\bot}=\omega r_{i}$. We also assume that the energy density of the
extremities of the flux tube is modified of a negative constant $C/2$,
in order to take into account possible boundary effects due to the
contact between the tube and the quark. Further, we consider that the
quarks can interact via $V(r)$ taking into account a short-range
potential (a one-gluon-exchange process, for instance). These two extra
terms are discussed in Ref.~\cite{sema95}. The
Lagrangian ${\cal L}$ of the meson is given by
\begin{eqnarray}
{\cal L}&=&{\cal L}_{1}+{\cal L}_{2}-V(r), \label{glrft} \\
{\cal L}_{i}&=&-m_{i}\gamma^{-1}_{i}-a\int^{r_{i}}_{0}dr^{'}_{i}\,
\gamma^{'\, -1}_{i\bot}-\frac{C}{2}\gamma^{-1}_{i\bot},
\end{eqnarray}
where $m_{i}$ is the constituent mass of the $i$th quark,
$\gamma_{i}=(1-v^{2}_{i})^{-1/2}$ and
$\gamma_{i\bot}=(1-v^{2}_{i\bot})^{-1/2}$.

\par In the following, we will only consider the symmetric case,
$m_{1}=m_{2}\equiv m$. Then, $r_{1}=r_{2}$, $r=2r_{1}$,
$v_{1\bot}=v_{2\bot}= v_{\bot}$. The corresponding quantized
equations of the system are \cite{laco89,sema95}
\begin{eqnarray}
\label{tf_equa1}
\frac{2\sqrt{\ell(\ell+1)}}{r}&=&\{v_{\bot}\gamma_{\bot},W_{r}\}+a\{r,f(
v_{\bot})\}+Cv_{\bot}\gamma_{\bot}, \\
\label{tf_equa2}
H&=&\{\gamma_{\bot},W_{r}\}+\frac{a}{2}\left\{r,\frac{\arcsin
v_{\bot}}{v_{\bot}}\right\}+C\gamma_{\bot}+V(r),
\end{eqnarray}
where $\ell$ is the orbital angular momentum, $\{A, B\}=AB+BA$,
$4x^{2}f(x)=\arcsin x-x\sqrt{1-x^{2}}$, $W_{r}=\sqrt{p^{2}_{r}+m^{2}}$,
and $p^{2}_{r}\equiv -\frac{1}{r}\frac{\partial^{2}}{\partial r^{2}}r$.
The operator $v_{\bot}$ commutes neither with $r$ nor with $p_{r}$
operators. These equations reduce to a spinless Salpeter equation with
the potential $ar+V(r)+C$ when $\ell=0$, and to a Schr\"{o}dinger
equation with the same potential in the nonrelativistic limit. The
general case ($m_{1}\neq m_{2}$) is detailed in Ref.~\cite{olss95}.

\subsection{The rotating string}

Starting from the QCD Lagrangian and writing the gauge invariant
$q\bar{q}$ Green function for confined spinless quarks in the Feynman-
Schwinger representation, one can arrive at the Nambu-Goto Lagrangian,
which describes two quarks with masses $m_{1}$ and $m_{2}$, attached by
a string of energy density $a$. With the straight line ansatz and the
introduction of auxiliary fields (einbein fields) to get rid of the
square roots appearing in this Lagrangian, one can obtain the
Hamiltonian
\begin{equation}
\label{rs1}
H=\frac{1}{2}\left[\frac{p^{2}_{r}+m^{2}_{1}}{\mu_{1}}+\frac{p^{2}_{r}+m
^{2}_{2}}{\mu_{2}}+\mu_{1}+\mu_{2}+a^{2}r^{2}\int^{1}_{0}\frac{d\beta}{
\nu}+
\int^{1}_{0}d\beta\nu+\frac{L^{2}}{a_{3}r^{2} } \right] + V(r),
\end{equation}
where
\begin{equation}\label{a3def}
a_{3}=\mu_{1}(1-\zeta)^{2}+\mu_{2}\zeta^{2}+\int^{1}_{0} d\beta\,
(\beta-\zeta)^{2}\, \nu.
\end{equation}
The potential $V(r)$ takes into account interactions not simulated by
the
rotating string. We do not consider here a contribution coming from a
constant potential $C$, as in the RFT model. $L=\sqrt{\ell(\ell+1)}$ and
$\zeta$ defines the position $R_{\mu}$ of the center of mass:
$R_{\mu}=\zeta x_{1\mu}+(1-\zeta)x_{2\mu}$, where $x_{i\mu}$ is the
coordinate of the $i$th quark. The auxiliary fields $\mu_{1}$ and
$\mu_{2}$ can be seen as effective masses of the quarks, while the
auxiliary field $\nu$ can be interpreted as an effective energy
density for the string.

\par We are interested here in the resolution of the symmetrical case.
When $m_{1}=m_{2}= m$, then $\zeta = 1/2$ and
$\mu_{1}=\mu_{2}=\mu$. Defining
\begin{equation}
\label{ydef}
y=\frac{L}{2a_{3}r},
\end{equation}
one can eliminate $\nu$ by a variation of the Hamiltonian. This extremal
field $\nu_{0}$ reads
\begin{equation}
\nu_{0}=\frac{ar}{\sqrt{1-4y^{2}(\beta-1/2)^{2}}}.
\end{equation}
By replacing $\nu$ by $\nu_{0}$ in the Hamiltonian~(\ref{rs1}) and the
relation~(\ref{ydef}), we
obtain the following equations for the symmetrical
rotating string \cite{morg99}
\begin{eqnarray}
\label{rs2}
\frac{\sqrt{\ell(\ell+1)}}{ar^{2}}&=&\frac{\mu y}{ar}+\frac{1}{4y^{2}}
\left( \arcsin y-y\sqrt{1-y^{2}} \right), \\
\label{rs3}
H&=&\frac{p^{2}_{r}+m^{2}}{\mu}+\mu+\frac{ar}{y}\arcsin y+\mu
y^{2}+V(r).
\end{eqnarray}

It has been shown in Ref.~\cite{sema04} that the extremal value of
$\mu$ giving $\delta H/\delta \mu=0$ is
\begin{equation}
\label{muextr}
\mu_{0}=\sqrt{\frac{p^{2}_{r}+m^{2}}{1-y^{2}}}.
\end{equation}
Moreover, the replacement of $\mu$ by $\mu_{0}$ in Eqs.~(\ref{rs2})
and (\ref{rs3}) gives exactly the symmetrical RFT equations
(\ref{tf_equa1}) and (\ref{tf_equa2}), with $y$ equal to
$v_{\bot}$. The RS model with all its auxiliary fields eliminated is
thus equivalent to the RFT model in the classical
symmetrical case. This is also true when ($m_{1}\neq m_{2}$), as shown
in Ref.~\cite{buis04}.

Here, we use the RS model with the auxiliary field $\mu$ not
eliminated, as in Refs.~\cite{dubi94,morg99}. In these papers, the
parameter $\mu$ is considered as a real parameter and not as an
operator. But, to avoid
eventual singularities in the value of this auxiliary field when $y$ is
classically close to $1$, we introduce  explicitly the dependance of
$\mu$ in $y$,
through the following substitution
\begin{equation}
\label{substitu}
\mu\rightarrow\frac{\rho}{\sqrt{1-y^{2}}},
\end{equation}
where $\rho$ is a real number. Such an expression is inspired by the
result~(\ref{muextr}). The
quantized equations of the symmetrical rotating string are thus
\begin{eqnarray}
\label{rs4}
\frac{\sqrt{\ell(\ell+1)}}{r}
&=&\rho \frac{y}{\sqrt{1-y^{2}}}+
\frac{a}{2}\left\{r,f(y)\right\}, \\
\label{rs5}
H&=&\frac{1}{2\rho}\left\{p^{2}_{r}+m^{2},
\sqrt{1-y^{2}}\right\}+\rho\frac{1+y^{2}}{\sqrt{1-y^{2}}}
+\frac{a}{2}\left\{r,\frac{\arcsin y}{y}\right\}+V(r),
\end{eqnarray}
where $4x^{2}f(x)=\arcsin x-x\sqrt{1-x^{2}}$ like for the RFT model.

A particular solution depends on the value of this parameter $\rho$.
Following Refs.~\cite{dubi94,morg99}, the physical value of $\rho$
minimizes the mass of the state. The mean value
$\langle \mu \rangle = \langle \rho/\sqrt{1-y^{2}} \rangle$ can be
considered as a constituent mass for the quark, depending on the state.
These equations reduce to a Schr\"{o}dinger-like equation with
the potential $ar+V(r)$ when $\ell=0$ \cite{sema04}, and to a true
Schr\"{o}dinger
equation with the same potential in the nonrelativistic limit.

\section{Resolution}
\label{resolution}

\subsection{The relativistic flux tube}

The main purpose of our work is the resolution of the symmetrical flux
tube equations (\ref{tf_equa1}) and (\ref{tf_equa2}) using the
Lagrange mesh method. To do this, we have to compute the matrix elements
of the different operators in the Lagrange basis. As we consider a
radial problem, we will use a Gauss-Laguerre quadrature. So, the
corresponding Lagrange functions will be given by
Eq.~(\ref{flag}). Let us define the
different matrix elements we need to know
\begin{equation}\label{defprem}
\begin{array}{lll}
A_{ij}=\left\langle f_i\left|\frac{2\sqrt{\ell(\ell+1)}}{r}\right|f_j
\right\rangle  , &
B_{ij}=\left\langle f_i|r|f_j\right\rangle, &
D_{ij}=\left\langle f_i|W_{r}|f_j\right\rangle  , \\
F_{ij}=\left\langle f_i\left|\frac{\arcsin v_{\bot}}{4v^{2}_{\bot}}-
\frac{
\sqrt{1-v^{2}_{\bot}}}{4v_{\bot}}\right|f_j \right\rangle, &
G_{ij}=\left\langle f_i|v_{\bot}\gamma_{\bot}|f_j\right\rangle, &
S_{ij}=\left\langle f_i\left|\frac{\arcsin v_{\bot}}{v_{\bot}}\right|f_j
\right\rangle   , \\
\Gamma_{ij}=\left\langle f_i|\gamma_{\bot}|f_j\right\rangle , &
V_{ij}=\left\langle f_i|V(r)|f_j\right\rangle. &
\end{array}
\end{equation}
With these notations, Eqs.~(\ref{tf_equa1}) and (\ref{tf_equa2}) read
\begin{equation}\label{eqmatri1}
A=\{G,D\}+a\{B,F\}+CG   ,
\end{equation}
\begin{equation}\label{eqmatri2}
H=\{\Gamma,D\}+\frac{a}{2}\{B,S\}+C\Gamma+V,
\end{equation}
where we have used the approximative closure relation,
\begin{equation}
\sum^{N}_{k=1}|f_{i}\rangle\langle f_{j}|\approx \openone,
\end{equation}
to compute a product of two matrices.

The matrix elements $A_{ij}$, $B_{ij}$, and $V_{ij}$ are easy to
compute, thanks to Eq.~(\ref{poten}). Moreover,
Eq.~(\ref{pij_def}) gives us an analytical expression for
$p^{2}_{r\, ij}$,
from which we can deduce the matrix elements $D_{ij}$ by using the
procedure described in
Sec.~\ref{Meth}. The same procedure will allow us to compute $F_{ij}$,
$G_{ij}$, $S_{ij}$ and $\Gamma_{ij}$ once the matrix elements of
$v_{\bot}$ are known. The determination of these matrix elements can be
achieved by an iterative process, described here:
\begin{enumerate}
\item Equation~(\ref{eqmatri1}) can be rewritten as
\begin{equation}
\label{iter}
G=\frac{1}{2}\{P,D^{-1}\}-\frac{C}{2}\{G,D^{-1}\}-\frac{1}{2}DGD^{-1}-
\frac{1}{2}D^{-1}GD,
\end{equation}
where
\begin{equation}
P=A-a\{B,F\}.
\end{equation}
This equation is symmetrized to ensure that $G$ is
hermitian. It is worth noting that $P=P(G)$ since $F=F(G)$. Starting
from an known matrix $G^{k}$ at the $k$th step, $P^{k}$ can be
computed and we obtain a new matrix $G^{k'}$ with Eq.~(\ref{iter}).
\item This iterative process would diverge if we choose
$G^{k+1}=G^{k'}$. So, we introduce a new parameter $\epsilon<1$ and
define $G^{k+1}= \epsilon G^{k'}+(1-\epsilon) G^{k}$.
\item At each step $k$, the $N$ eigenvalues
$\left\{v_{\bot\, i}^{(k)}\right\}$ of
the operator $v_{\bot}$ are computed. The iteration procedure ends when
\begin{equation}
\frac{1}{N} \sum_{i=1}^N \left| \frac{v_{\bot\, i}^{(k+1)} -v_{\bot\, i}
^{(k)}}{v_{\bot\, i}^{(k+1)}} \right| < \eta,
\end{equation}
where $\eta$ is a fixed tolerance.
\end{enumerate}
Once we have reached the convergence for $G$, we are able to compute $S$
and $\Gamma$, which are now seen as functions of the matrix $G$ rather
than the matrix elements of the operator $v_{\bot}$. The Hamiltonian can
then be computed and diagonalized.

\par Actually, the final matrix $G$ is practically independent of the
initial one $G^{0}$. However, the faster way to reach the convergence is
to develop Eq.~(\ref{tf_equa1}) at the first order in $v_{\bot}$ and
to choose the matrix $G$ given by this development. At the first
order, $v_{\bot}\gamma_{\bot}\approx v_{\bot}$, and
\begin{equation}
G^{0} \approx \sqrt{\ell(\ell+1)}\left(
\frac{1}{2}\{B,D\}+\frac{aB^{2}}{6}+\frac{C}{2}\openone \right)^{-1}.
\end{equation}
Let us note that a relevant starting matrix is obtained even if $m=0$.

\subsection{Rotating string}
\label{Rota}

\subsubsection{Lagrange mesh method}

The resolution of the RS with the Lagrange mesh method is similar to
that of the RFT. Indeed, using the previous definitions~(\ref{defprem})
with $y$ instead of $v_{\bot}$, and defining
\begin{equation}
\begin{array}{lll}
Q_{ij}=\left\langle f_i\left|\sqrt{1-y^{2}} \right| f_j \right\rangle
, &
Y_{ij}=\left\langle f_i\left|\frac{1+y^{2}}{\sqrt{1-y^{2}}} \right| f_j
\right\rangle   , &
E_{ij}=\left\langle f_i\left|p_r^2+m^2 \right| f_j \right\rangle ,
\end{array}
\end{equation}
Eqs.~(\ref{rs4}) and (\ref{rs5}) are given by
\begin{eqnarray}
\label{co_t_g}
G&=&\frac{1}{2\rho}\left(A-a\{B,F\} \right)  ,\\
\label{co_t_h}
H&=&\frac{1}{2\rho}\{E,Q\}+\rho Y+\frac{a}{2}\{B,S\}+V.
\end{eqnarray}
Like for the RFT, we need to compute the matrix of the operator $y$ to
completely know the Hamiltonian. We will do this by an iterative
process on $G$ given directly by Eq.~(\ref{co_t_g}), with an initial
value, obtained after a first order
development, given by
\begin{equation}
G^{0}=\sqrt{\ell(\ell+1)}\left(\rho B+\frac{a}{6}B^{2}\right)^{-1}.
\end{equation}
The last step in the resolution of the RS equations is always
to find the value of the real number $\rho$ realizing the minimum mass
of a particular state. This extremal value is different for each state.

\subsubsection{WKB method}

Contrary to the case of the RFT, the operator $p^{2}_{r}$ appears only
in the equation defining the Hamiltonian for the RS. This makes possible
a solution of Eqs.~(\ref{rs4}) and (\ref{rs5}) by a WKB method.

First, let us examine the case $\ell=0$. The RS equations reduce then
to a spinless Salpeter equation of the form ($\rho=\mu$ since $y=0$)
\begin{equation}
H=\frac{\vec{p}^{\, 2}+m^{2}}{\rho}+\rho+ar+V(r),
\end{equation}
where
\begin{equation}
\vec{p}^{\,2}=p^{2}_{r}+\frac{L^{2}}{r^{2}}.
\end{equation}
In the WKB method, $L=\ell+1/2$. Consequently $L^{2}=1/4$ here, and we
obtain
\begin{equation}
p^{2}_{r}=\rho M-\rho^{2}-m^{2}-\rho ar-\frac{1}{4r^{2}}-\rho V(r),
\end{equation}
where $M$ is the meson mass.
We have then to compute $r_{+}$ and $r_{-}$ the two physical zeros of
the classical quantity
$p^{2}_{r}$. Finally, the resolution of the Bohr-Sommerfeld condition
\begin{equation}
\label{wkb_1}
\int^{r_{+}}_{r_{-}} p_{r}\, dr= \pi\left(n+\frac{1}{2}\right),
\end{equation}
followed by a minimization of $M$ with respect to the parameter $\rho$
gives the mass of the state whose quantum numbers are $\ell$ and $n$.

When $\ell\neq0$, the WKB formulation of the classical RS
equations~(\ref{rs2}) and (\ref{rs3}), with the
substitution~(\ref{substitu}), reads
\begin{eqnarray}
\label{semic_1}
\frac{\ell+\frac{1}{2}}{ar^{2}}&=&\frac{\rho
y}{ar\sqrt{1-y^{2}}}+\frac{1}{4y^{2}}\left(\arcsin y-y\sqrt{1-y^{2}}
\right), \\
M&=&\frac{1}{\rho}\left(p^{2}_{r}+m^{2}\right)\sqrt{1-y^{2}}+\rho\frac{
1+y^{2}}{\sqrt{1-y^{2}}}+\frac{ar}{y}\arcsin y+V(r).
\end{eqnarray}
The first one implicitly defines a function $y=\tilde{y}(r,\ell,\rho)$,
which can be numerically computed. We can then
formally write
\begin{equation}
p^{2}_{r}=\frac{\rho}{\sqrt{1-\tilde{y}^{2}}}(M-V(r)-\frac{ar}{\tilde{y
}}\arcsin
\tilde{y})-\rho^{2}\frac{1+\tilde{y}^{2}}{1-\tilde{y}^{2}}-m^{2}.
\end{equation}
The rest of the resolution is now identical to the previous case
$\ell=0$.

\section{Set of parameters}\label{params}
\subsection{Physical parameters}

In this paper, we are mainly interested in the capacity of our method to
give accurate solutions of the coupled equations for both RFT and RS
models. But, in order to compare our results with previous studies and
to use our method with physical parameters in interesting ranges, we
will use the values of physical quantities from the models Ia and Ic
developed in Ref.~\cite{sema95} (see Table~\ref{tab:sets}). Both models
possess a coulomb term with three values of the strength, depending on
the quark content of the meson: $\kappa_{hl}$ for heavy-light system,
$\kappa_{hh}$ for heavy-heavy system, and $\kappa_{ll}$ for light-light
system (light quark: $u$, $d$, $s$; heavy quark: $c$, $b$).

\subsection{The scale parameter}

The Lagrange mesh method provides us a direct picture of the wave
function at the mesh points. The best results are thus obtained when the
mesh covers the main part of the wave function and the last mesh point
is located in the asymptotic tail. That is why we are interested in an
adequate determination of the scale parameter $h$. Since the method is
not
variational, no extremum of the mass can be expected for a defined value
of $h$. A good value for this quantity is given by $h=r_{a}/x_{N}$,
where $x_{N}$ is $N$th zero of the Laguerre polynomial (the last point
of the mesh), and $r_{a}$ represents a distance for which the asymptotic
tail of the wave function is well defined. If $x_{N}$ is well known,
$r_{a}$ is not. We show here how such a quantity can be estimated.

A typical evolution of the computed masses for different values of $h$
is presented in  Fig.~\ref{fig:hevol}. The existence of plateaus shows
that the method does not require the knowledge of precise values of the
scale parameter. A simple estimation will be sufficient, even to obtain
accurate results.

\par For given quantum numbers, a system of two massless quarks is
expected
to have the maximal spatial extension, and so it could give an upper
bound of the parameter $h$. First, we analyze the problem for the RS
equations when $\ell=0$. These equations reduce then to a spinless
Salpeter Hamiltonian, which reads
\begin{equation}
H_{A}=\frac{\vec{p}^{\,2}}{\rho}+\rho+ar.
\end{equation}
We fix $V(r)=0$, since the asymptotic behavior is controlled by the
confinement.
The solutions have the following analytical forms ($n=$ 0, 1, \ldots)
\cite{sema04}
\begin{eqnarray}
E_{n0}(\rho)&=&\left(\frac{a^{2}}{\rho}\right)^{1/3} (-s_{n})+\rho,
\\
u_{n0}(r)&=&(\rho a)^{1/6} \frac{Ai\left((\rho a)^{1/3}r+s_{n}\right)}{|
Ai'(s_{n})|},
\end{eqnarray}
where $Ai(s)$ is the Airy function and $s_{n}$ its $n$th zero, given by
the approximate formula \cite{abra70}
\begin{equation}\label{zero_ai}
s_{n}\approx-\left[\frac{3\pi}{2}(n+\frac{3}{4})\right]^{2/3}.
\end{equation}
Replacing $\rho$ by its extremal value $\rho_{n0}$,
\begin{equation}
\rho_{n0}=\sqrt{a}\left(\frac{-s_{n}}{3}\right)^{3/4},
\end{equation}
we have
\begin{equation}
u_{n0}(r)\div Ai\left(\sqrt{a}\left(\frac{-s_{n}}{3}\right)^{1/4}r+s_{n}
\right).
\end{equation}
When $s\approx5$, $Ai(s)$ is about $0.02\%$ of its maximal value.
Consequently, a good estimation of $r_{a}$ is given by
\begin{equation}
\sqrt{a}\left(\frac{-s_{n}}{3}\right)^{1/4}r_{a}+s_{n}=5.
\end{equation}

At this point, we are able to compute a ``physical" value for $h$
when $\ell=0$. The extension of the wave function
increases with the angular momentum. The simplest way to simulate such
an increase is to compute $r_{a}$ with the relation
\begin{equation}
\label{hdef}
\sqrt{a}\left(\frac{-s_{n+\ell}}{3}\right)^{1/4}r_{a}+s_{n+\ell}=5.
\end{equation}
This crude estimation of $h$ is satisfactory because it is always
located in the plateau. Moreover, we will use it in both RFT and RS
methods, because of the classical equivalence between these two
theories.

\subsection{Numerical parameters}

The accuracy of the solutions depends mainly on two parameters: the
number $N$ of mesh points (basis states) and the value of the tolerance
$\eta$ on the eigenvalues of the operator $v_{\bot}$. For instance, a
relative error on meson masses around $10^{-5}$ can be reached with
$N \ge 30$ and $\eta \le 10^{-6}$. The accuracy can be increased by
using greater values of $N$ and smaller values of $\eta$.

If the value of the mixing parameters $\epsilon$ is too high, the
iterative process diverges. The best value of $\epsilon$ is chosen as
the largest value for which the process converges. It depends on the
case considered, as shown in Table~\ref{tab:epsilon}. It clearly appears
that the iterative process does not converge easily with the RFT
equations, especially when the quarks are massless. About $700$
iterations are needed in this case, and $400$ when
$m/\sqrt{a} \gtrsim 1$. However, the RS solutions converge faster, and
one can reach the convergence after about only $40$ iterations.

\section{Results}\label{Resu}

\subsection{Relativistic flux tube}
\label{resuRFT}

We have computed with the Lagrange mesh method the solutions of the RFT
equations for models Ia and Ic from Ref.~\cite{sema95} (see
Table~\ref{tab:sets}). All the masses are computed with $N=30$,
$\eta=10^{-6}$, the scale parameter $h$ is estimated thanks to
Eq.~(\ref{hdef}), and the parameter $\epsilon$ is taken from
Table~\ref{tab:epsilon}. Meson masses are presented in
Table~\ref{tab:resu1} with the corresponding ones computed with the
method developed in Ref.~\cite{sema95}, relying on a harmonic oscillator
basis. Experimental data are given in order to show that the parameters
used are physically relevant.

The results of both methods are compatible. Nevertheless, the masses
computed with the Lagrange mesh method are always smaller than the
masses computed with the harmonic oscillator method, although the
Lagrange mesh method is not variational. Our method provides thus a
better convergence of the results. The improvement is especially
important for light quark masses. Differences between the two methods
vanish when the quark mass increases.

It is worth noting that the masses computed with method of
Ref.~\cite{ sema95} are strongly dependent of the values chosen for the
oscillator length. So a supplementary minimization on this
parameter, for each state, is necessary to obtain the optimal value of
a mass. This is not necessary with the Lagrange mesh method since it
is nearly independent of the scale parameter (see Fig.~\ref{fig:hevol}).

The small differences between the masses obtained with the Lagrange mesh
method and the harmonic oscillator method are a strong indication that
our method works well. But we want another test. It will be given by the
study of the Rotating string model.

\subsection{Rotating string}
\label{resuRS}

Solutions of the RS equations computed with the Lagrange mesh method
(numerical parameters as in Sec.~\ref{resuRFT}) and the WKB
approximation are presented in Table~\ref{tab:resu2}. The masses are
obtained using the set Ia of parameters (see Table~\ref{tab:sets}), for
a pure string without coulomb-like potential.

The two methods to solve the RS equations lead to very close results.
This shows that the semiclassical approximation is efficient in this
case, but also that the Lagrange mesh method works correctly.
Fig.~\ref{fig:minimum} shows the existence of a minimal mass for a
particular value $\rho_0$ of the parameter $\rho$ in the RS equations.
In our calculations, $\rho_{0}$ has been determined to the nearest
$10$~MeV, and is the same in the two methods with that precision. An
accuracy below 1~MeV is then reached for the masses.

\subsection{Comparison between RFT and RS}

If the RS model is classically equivalent to the RFT model once the
auxiliary fields are correctly eliminated, the two models should not
give the same results when a real parameter $\rho$ is kept in the RS
equations. In a previous study \cite{sema04}, some results have been
obtained about the equivalence between a spinless Salpeter Hamiltonian
$H_{SS}$ and a corresponding Hamiltonian with auxiliary field $H_A$: the
eigenvalues of $H_A$ are upper bound of the eigenvalues of $H_{SS}$
\cite{luch96} with relative  differences around 7\% for the lowest
states. We know that the RFT and RS equations reduce respectively to
eigenvalues equations for Hamiltonians $H_{SS}$ and $H_A$ for a
vanishing angular momentum. It should be interesting to see if there is
the same kind of relation between the masses for the RFT and RS models
when $\ell \ne 0$.

Another result can be expected: once we know an eigenfunction
$|\psi_{RS}\rangle$ of the RS Hamiltonian for the extremal value
$\rho_{0}$, we are able to compute the effective mass
$\mu_{0}=\rho_{0}\langle 1/\sqrt{1-y^{2}}\, \rangle$ for this state.
This quantity should be approximately equal to the mean value
$\mu_{RFT}=\left\langle \sqrt{(p^{2}_{r}+m^{2})/(1-v^{2}_{\bot})}\,
\right\rangle$ for the corresponding state $|\psi_{RFT}\rangle$, due to
the equivalence between the two models via Eq.~(\ref{muextr}).

Our results are given in Table~\ref{tab:resu3}. The masses for both RFT
and RS models are computed with the Lagrange mesh method for the same
parameters as in Sec.~\ref{resuRS}. The RS masses are always upper bound
of the RFT masses with relative differences around by 7\%, as in the
limiting case of vanishing angular momentum. We also see that
$\mu_{RFT}\approx\mu_{0}$ as expected. We can finally notice that the
results of the two models are closer and closer when the mass of the
constituent quark increases, because the RFT and the RS model posses and
common nonrelativistic limit: the Schr\"{o}dinger equation with a linear
potential.

\section{Conclusions}
\label{conclu}

We have shown in this paper that the Lagrange mesh method solves
successfully the equations of the relativistic flux tube model in the
symmetrical case. The masses obtained are in good agreement with a
previous resolution in a harmonic oscillator basis \cite{sema95}. But
the Lagrange mesh method is more efficient, due to its independence of
the scale parameter used to fit the size of the trial states. Moreover,
a better convergence is reached. This proves the validity of our method.

We have also solved the equations of the symmetrical rotating string
model with the Lagrange mesh method and with the WKB approximation. The
masses computed with these two procedures are very close, showing that
the Lagrange mesh method correctly works, and that the WKB approximation
is efficient here. If we compare the masses given by the relativistic
flux tube and the rotating string models, we find relative differences
around $7$\% for the lowest states, as expected because the two models
are classically equivalent. This point is a last confirmation of the
relevance of the Lagrange mesh method to solve the relativistic flux
tube equations.

\acknowledgments

C.~S. (FNRS Research Associate) and F.~B. (FNRS Research
Fellow) thank the FNRS for financial support.

\clearpage

\begin{table}[ht]
\protect\caption{Two sets of physical parameters for the RFT and the RS
models, from Ref.~\cite{sema95} ($n$ = $u$ or $d$).}
\label{tab:sets}
\begin{ruledtabular}
\begin{tabular}{lll}
 & Ia & Ic\\
\hline
$m_{n}$ (GeV) & 0 & 0.233\\
$m_{s}$ (GeV) & 0.317 & 0.416\\
$m_{c}$ (GeV) & 1.456 & 1.658\\
$a$  (GeV$^{2}$) & 0.151 & 0.169\\
$C$ (GeV) & 0 & $-2m_{n}$\\
$\kappa_{ll}$ & 1.016 & 0.539\\
$\kappa_{hl}$ & 0.698 & 0.467\\
$\kappa_{hh}$ & 0.544 & 0.500 \\
\end{tabular}
\end{ruledtabular}
\end{table}

\begin{table}[h]
\protect\caption{Approximate optimal values for the parameter $\epsilon$
in different cases.}
\label{tab:epsilon}
\begin{ruledtabular}
\begin{tabular}{lll}
 & \multicolumn{2}{c}{$\epsilon$} \\
\cline{2-3}
$m/\sqrt{a}$ & RFT & RS \\
\hline
$\approx 0$  & $0.005$ & $0.1$ \\
$\gtrsim 1$ & $0.01$  & $0.1$ \\
\end{tabular}
\end{ruledtabular}
\end{table}

\begin{table}[ht]
\protect\caption{Meson masses for the RFT model, with two sets Ia  and
Ic of parameters from Ref.~\cite{sema95}, computed using the Lagrange
mesh method (Lag.) and a previous technique relying on an harmonic
oscillator basis (HO) \cite{sema95}. The experimental masses (Exp.) are
given, without error, for information.}
\label{tab:resu1}
\setlength{\extrarowheight}{1pt}
\begin{ruledtabular}
\begin{tabular}{lllllll}
&  & \multicolumn{5}{c}{Mass (GeV)}\\
\cline{3-7}
 & $(n+1)^{2S+1}L_{J}$ & Exp. & HO (Ia)  & Lag.(Ia) & HO (Ic)  &
 Lag. (Ic)\\
\hline\hline
$n\bar{n}$ & $1 ^{3}S_{1}$ & 0.771 & 0.781 & 0.762 & 0.774 & 0.773 \\
& $1 ^{3}P_{2}$ & 1.318 & 1.310 & 1.300 & 1.320 & 1.319 \\
& $1 ^{3}D_{3}$ & 1.691 & 1.654 & 1.643 & 1.689 & 1.676 \\
& $2 ^{3}S_{1}$ & 1.465 & 1.450 & 1.415 & 1.427 & 1.424\\
& $2 ^{3}P_{2}$ & 1.810 & 1.841 & 1.832 & 1.797 & 1.794 \\
\hline
$s\bar{s}$& $1 ^{3}S_{1}$ & 1.019 & 0.988 & 0.968 & 1.010 & 1.010 \\
& $1 ^{3}P_{2}$ & 1.525 & 1.540 & 1.534 & 1.517 & 1.515\\
& $1 ^{3}D_{1}$ & 1.854 & 1.881 & 1.877 & 1.867 & 1.865 \\
& $2 ^{3}S_{1}$ & 1.680 & 1.671 & 1.641 & 1.644 & 1.641\\
& $2 ^{3}P_{2}$ & 2.011 & 2.053 & 2.047 & 1.994 & 1.991\\
\hline
$c\bar{c}$ & $1 ^{3}S_{1}$ & 3.097 & 3.131 & 3.130 & 3.116 & 3.115\\
& $1 ^{3}P_{2}$ & 3.556 & 3.528 & 3.527 & 3.542 & 3.542\\
& $1 ^{3}D_{3}$ & 3.770 & 3.788 & 3.788 & 3.820 & 3.820\\
& $2 ^{3}S_{1}$ & 3.686 & 3.666 & 3.663 & 3.664 & 3.661\\
& $2 ^{3}D_{1}$ & 4.159 & 4.128 & 4.128 & 4.165 & 4.164\\
%\hline
%$b\bar{b}$ & $1 ^{3}S_{1} $ & 9.460 & 9.447 & 9.442 & 9.454& 9.450\\
%& $1 ^{3}P_{2}$ & 9.913 & 9.948 & 9.950 & 9.940 & 9.943\\
%& $2 ^{3}S_{1}$ & 10.023 & 10.023 & 10.019 & 10.017 & 10.014\\
%& $2 ^{3}P_{2}$ & 10.269 & 10.270 & 10.271 & 10.270 & 10.272\\
\end{tabular}
\end{ruledtabular}
\end{table}

\begin{table}[ht]
\protect\caption{Meson masses for the RS model computed with the
Lagrange mesh method and the WKB approximation. The interaction Ia from
Ref.~\cite{sema95} is used, but without the Coulomb potential. The
extremal value $\rho_0$, to the nearest 10~MeV, of
the parameter $\rho$ is given in both cases.}
\label{tab:resu2}
\setlength{\extrarowheight}{1pt}
\begin{ruledtabular}
\begin{tabular}{llllll}
 & &\multicolumn{2}{c}{Lagrange mesh}&\multicolumn{2}{c}{WKB}\\
\cline{3-6}
 & & Mass (GeV) & $\rho_{0}$ (GeV) & Mass (GeV) &
 $\rho_{0}$ (GeV) \\
\hline
$n\bar{n}$ & $1^{3}S_{1}$ & 1.289 & 0.32 & 1.294 & 0.32 \\
           & $1^{3}P_{2}$ & 1.581 & 0.16 & 1.589 & 0.16 \\
           & $2^{3}S_{1}$ & 1.960 & 0.49 & 1.963 & 0.49 \\
\hline
$c\bar{c}$ & $1^{3}S_{1}$ & 3.492 & 1.55 & 3.493 & 1.55 \\
           & $1^{3}P_{2}$ & 3.731 & 1.52 & 3.730 & 1.51 \\
           & $2^{3}S_{1}$ & 3.916 & 1.65 & 3.917 & 1.65 \\
\end{tabular}
\end{ruledtabular}
\end{table}

\begin{table}[ht]
\protect\caption{Meson masses computed with the
Lagrange mesh method within the RFT and the RS models. The interaction
Ia from Ref.~\cite{sema95} is used, but without the Coulomb potential.
The values of the effective masses $\mu_0$ and $\mu_{RFT}$ are given.}
\label{tab:resu3}
\begin{ruledtabular}
\begin{tabular}{llllll}
 & & \multicolumn{2}{c}{Relativistic flux tube}& \multicolumn{2}{c}{
 Rotating string}\\
\cline{3-6}
 & & Mass (GeV) & $\mu_{RFT}$ (GeV) & Mass (GeV) & $\mu_0$ (GeV)\\
\hline
$n\bar{n}$ & $1^{3}S_{1}$ & 1.228 & 0.308 & 1.289 & 0.32 \\
           & $1^{3}P_{2}$ & 1.543 & 0.323 & 1.581 & 0.29 \\
           & $1^{3}D_{3}$ & 1.825 & 0.342 & 1.860 & 0.32 \\
           & $2^{3}S_{1}$ & 1.832 & 0.460 & 1.960 & 0.49 \\
           & $2^{3}P_{2}$ & 2.071 & 0.498 & 2.155 & 0.49 \\
\hline
$s\bar{s}$ & $1^{3}S_{1}$ & 1.507 & 0.486 & 1.536 & 0.49 \\
           & $1^{3}P_{2}$ & 1.809 & 0.523 & 1.838 & 0.52 \\
           & $1^{3}D_{3}$ & 2.078 & 0.593 & 2.103 & 0.56 \\
           & $2^{3}S_{1}$ & 2.070 & 0.612 & 2.142 & 0.62 \\
           & $2^{3}P_{2}$ & 2.294 & 0.647 & 2.343 & 0.64 \\
\hline
$c\bar{c}$ & $1^{3}S_{1}$ & 3.486 & 1.555 & 3.492 & 1.55 \\
           & $1^{3}P_{2}$ & 3.723 & 1.594 & 3.731 & 1.58 \\
           & $1^{3}D_{3}$ & 3.931 & 1.625 & 3.937 & 1.62 \\
           & $2^{3}S_{1}$ & 3.902 & 1.631 & 3.916 & 1.65 \\
           & $2^{3}P_{2}$ & 4.081 & 1.661 & 4.094 & 1.66 \\
\end{tabular}
\end{ruledtabular}
\end{table}

\clearpage

\ \\
\begin{figure}[h]
% \includegraphics[0,0][11.65cm,9cm]{m_vs_h.bmp}
% 1650 x 1275
%\centerbmp{11.65cm}{9cm}{m_vs_h.bmp}
\caption{Typical evolution of meson masses for the RFT model with the
scale parameter $h$: $1S$, $1P$ and $2S$ states for the isospin 1
mesons,
computed with the parameters Ia from Ref.~\cite{sema95}.
Formula~(\ref{hdef}) gives $h=0.21$ GeV$^{-1}$ for the 1S state; this
value is correctly located in the plateau.}
\label{fig:hevol}
\end{figure}

\begin{figure}[htb]
% \includegraphics[0,0][11.65cm,9cm]{mu_min_art.bmp}
% 1650 x 1275
%\centerbmp{11.65cm}{9cm}{mu_min_art.bmp}
\caption{Meson masses for the RS model, computed with the Lagrange mesh
method, versus the parameter $\rho$: $1S$, $1P$ and $1D$ states for the
$s\bar{s}$ system, computed with the interaction Ia from
Ref.~\cite{sema95}, but without the Coulomb potential.}
\label{fig:minimum}
\end{figure}


\begin{thebibliography}{99}

\bibitem{baye86} D. Baye and P.-H. Heenen, J. Phys. A \textbf{19}, 2041
(1986).
\bibitem{vinc93} M. Vincke, L. Malegat, and D. Baye, J. Phys. B
\textbf{26}, 811 (1993).
\bibitem{baye95} D. Baye, J. Phys. B \textbf{28}, 4399 (1995).
\bibitem{hess99} M. Hesse and D. Baye, J. Phys. B \textbf{32}, 5605
(1999).
\bibitem{sem01} C. Semay, D. Baye, M. Hesse, and B. Silvestre-Brac,
Phys. Rev. E \textbf{64}, 016703 (2001).
\bibitem{laco89} D. LaCourse and M. G. Olsson, Phys. Rev. D \textbf{39},
2751 (1989).
\bibitem{olss95} M. G. Olsson and Sini\v{s}a Veseli, Phys. Rev. D
\textbf{51}, 3578 (1995).
\bibitem{sema95} C. Semay and B. Silvestre-Brac, Phys. Rev. D
\textbf{52}, 6553 (1995).
\bibitem{dubi94} A. Yu. Dubin, A. B. Kaidalov, and Yu. A. Simonov, Phys.
Lett. B \textbf{323}, 41 (1994).
\bibitem{morg99} V. L. Morgunov, A. V. Nefediev, and Yu. A. Simonov,
Phys. Lett. B \textbf{459}, 653 (1999).
\bibitem{sema04} C. Semay, B. Silvestre-Brac, and I. M. Narodetskii,
Phys. Rev. D \textbf{69}, 014003 (2004).
\bibitem{buis04} F. Buisseret and C. Semay, \emph{Auxiliary fields and
the flux tube model} [hep-ph/0406216].
\bibitem{baye02} D. Baye, M. Hesse, and M. Vincke, Phys. Rev. E
\textbf{65}, 026701 (2002).
\bibitem{abra70} M. Abramowitz and I. A. Stegun, \textit{Handbook of
mathematical functions} (Dover, New York, 1970).
\bibitem{luch96} W. Lucha and F. F. Sch\"{o}berl, Phys. Rev. A
\textbf{54}, 3790 (1996).

\end{thebibliography}
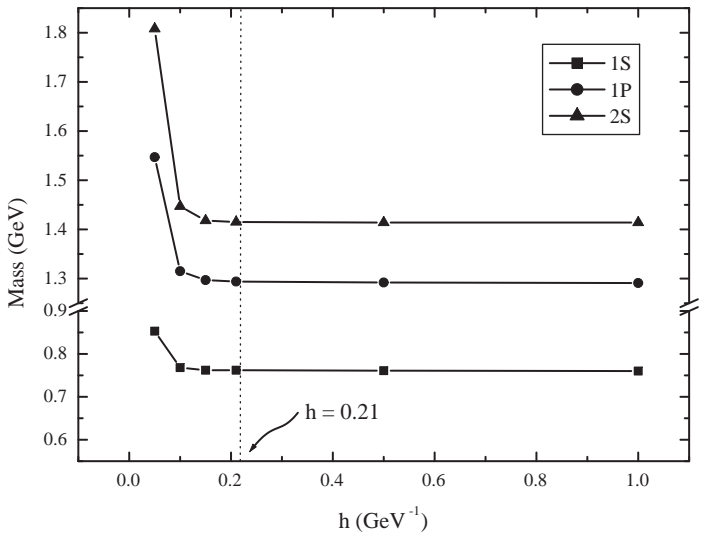
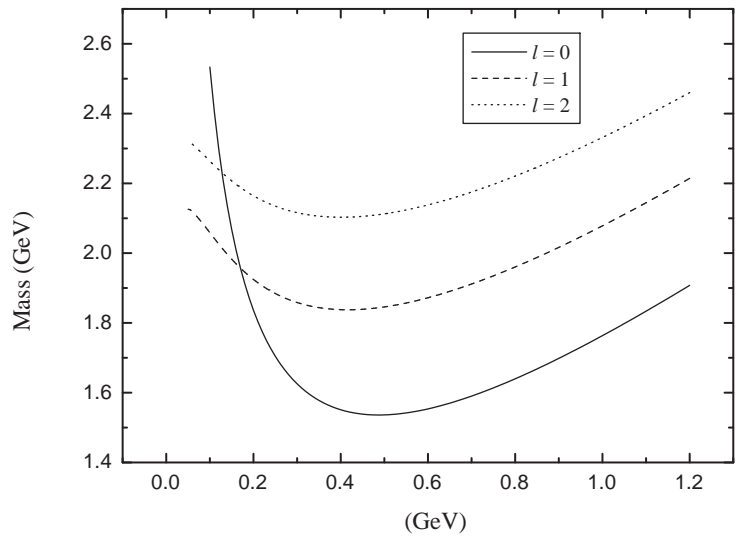
\end{document}